\documentclass[11pt]{article}

\newcommand{\BE}{\begin{equation}}
\newcommand{\EE}{\end{equation}}
\newcommand{\BA}{\begin{eqnarray}}
\newcommand{\EA}{\end{eqnarray}}

\begin{document}
\begin{titlepage}

\vspace*{1mm}
\begin{center}

            {\LARGE{\bf A gap-less mode of the singlet Higgs field } }

\vspace*{14mm}
{\Large  M. Consoli }
\vspace*{4mm}\\
{\large
Istituto Nazionale di Fisica Nucleare, Sezione di Catania \\
Corso Italia 57, 95129 Catania, Italy}
\end{center}
\begin{center}
{\bf Abstract}
\end{center}

Recent lattice results suggest the existence of a gap-less mode of the
singlet Higgs field.
We present a description of spontaneous symmetry breaking in 
$\lambda\Phi^4$ theories showing why
one is faced with long-wavelength,
collective modes of the scalar condensate whose energy
$\tilde{E}({\bf{p}}) \to 0$ 
in the ${\bf{p}} \to 0$ limit.
\end{titlepage}

\setcounter{equation}{0}
\section{Introduction}

Following analogies with condensed matter physics, 
the idea of a `condensed vacuum' is also playing 
an increasingly important role in
particle physics. Indeed, scalar, gluon and fermion condensates are basic 
ingredients in the present description of electroweak and strong 
interactions where one introduces
a set of elementary quanta whose perturbative vacuum state $|o\rangle$
is not the physical ground state $|v\rangle$ of the interacting theory. 

As pointed out in ref.\cite{salehi}, the issue of vacuum condensation 
in Lorentz-invariant theories is
somewhat controversial, in view of the apparent contradiction between 
exact Lorentz-invariance and a non-vanishing energy-density for the vacuum. 
For this reason, it has been suggested  that
the origin of vacuum condensation may
reflect a `principal violation' \cite{salehi}
of Lorentz-invariance, as in the case of the
existence of a fundamental length scale associated with an ultimate ultraviolet
 cutoff $\Lambda$. 

This remark is even more relevant if one attempts to
model the world as a sequence of effective theories embedded one
into the other. In this approach, the problem 
of exact Lorentz-covariance in an intermediate description is
unavoidable since the cutoff will never be really sent to infinity. However, 
even in a fully continuum, quantum-field theoretical approach \cite{soff} to 
$(\lambda\Phi^4)_4$ theory, where the vacuum is found to have exactly 
zero energy-density, one finds 
a non-trivial distribution function $f( {\bf{p}}^2 ) 
\equiv \langle N_{ {\bf{p}} }\rangle$ of the elementary quanta
in the non-perturbative vacuum. For this
reason, in the 
${\bf{p}} \to 0$ limit, where the $f( {\bf{p}}^2 )$ function
yields significant contributions, the propagation in the
physical ground state resembles the one in a condensed medium \cite{soff}.

Here we shall adopt the point of view that Lorentz-covariance has 
to become exact in the double limit where both the quantization
volume {\it and} the
ultraviolet cutoff are sent to infinity. However, for finite $\Lambda$ one 
may be faced with deviations from an exact Lorentz-covariant energy spectrum
that depend, in general, on the peculiar nature of the ground state.
In this sense, the formation of a `condensed vacuum', defined by
a macroscopic occupation of the same quantum state, may represent the operative
construction of a preferred reference frame 
suggesting non-Lorentz-covariant effects associated with
the {\it long-wavelength} excitations of the condensed phase.

To understand the underlying mechanisms, in the
case of spontaneously broken scalar theories,
we observe preliminarly that for an interacting system
the condensed vacuum cannot be a pure Bose condensate. Other components
are needed from purely phase-space arguments 
and take into account all possible processes that, with the same 
average particle density, can `deplete' the 
${\bf{p}}=0$ state. 

We shall assume that condensation extends 
up to some typical momentum 
$|{\bf{p}}| \sim \delta$ and is
controlled by an unknown 
distribution function 
$f_{\delta}( {\bf{p}}^2 )$
of the elementary quanta. It is clear that
$\delta \to 0 $ in the 
limit of a vanishingly small interaction where
$f_{\delta}( {\bf{p}}^2 )$ becomes non zero only for $ {\bf{p}}=0$. Therefore,
in the `trivial' \cite{book}
$\lambda\Phi^4$ theories, we expect the pure Bose condensate
to become a better and better approximation to the spontaneously broken 
vacuum when approaching the continuum limit of quantum field theory.
 
Let us now consider the excitation spectrum 
$\tilde{E}_\Lambda({\bf{p}})$ of the cutoff theory
in the limit when
$f_{\delta}( {\bf{k}}^2 )$ is very strongly peaked at 
${\bf{k}}=0$ and vanishes very quickly for
$|{\bf{k}}| \neq 0$. In this case, we can easily
understand the crucial difference 
with the usual perturbative vacuum. There, the low-lying excitations 
correspond to single particle states with $\sqrt{ {\bf{p}}^2 + m^2}$ 
energy and one has to reach 
$|{\bf{p}}| \sim m$ values
before two-particle excitations can become important. 
Here, in the condensed vacuum, for
$|{\bf{p}}| < \delta$ it is energetically 
more convenient to split the momentum among
a large number of quanta. 
This costs very little energy analogously to the small
displacements of atoms from their equilibrium positions, 
at the base of the propagation of elastic waves in continuous
media. On the other hand, for momenta larger than $\delta$, where 
the condensed vacuum looks `empty', a Lorentz-covariant spectrum applies, 
say $ \sqrt{ {\bf{p}}^2 + M^2_h}$. Since
$\delta \to 0$ when $\Lambda \to \infty$, the form
$\sqrt{ {\bf{p}}^2 + M^2_h}$  would hold everywhere 
in a strictly continuum theory but, at finite $\Lambda$, one has deviations 
from exact Lorentz-covariance in an infinitesimal region of momenta 
$|{\bf{p}}| < \delta$ (or equivalently for fluctuations with
wavelengths  larger than $\delta^{-1}$).

The possibility that the `Higgs mass'
$M_h$ differs non-trivially from the energy-gap $\tilde{E}_{\Lambda}(0)$ of
the broken phase
has been objectively addressed \cite{cea2} with lattice simulations. To this
end the energy spectrum was measured from the exponential decay
of the connected correlator at various values of $|{\bf{p}}|$. 
In this analysis, by approaching the continuum limit, one expects to find 
a region of 3-momentum where indeed $\tilde{E}_{\Lambda}({\bf{p}})$ is well 
reproduced by the asymptotic form 
$\sqrt{ {\bf{p}}^2 + M^2_h}$. The delicate issue is
whether this region extends 
down to  ${\bf{p}}=0$ and whether the value of
$M_h$ determined in this way agrees with the value of the energy gap
$\tilde{E}_{\Lambda}(0)$. To exclude possible trivial effects, 
one should also repeat the same analysis in the symmetric phase and 
compare the corresponding measured energy spectrum 
$E_{\Lambda}({\bf{p}})$ with the form
$\sqrt{ {\bf{p}}^2 + m^2}$. In the symmetric phase
the lattice data \cite{cea2}
give $E_{\Lambda}(0)=m$ to very high accuracy as expected.
However, in the broken phase, 
$\tilde{E}^2_{\Lambda}({\bf{p}})- {\bf{p}}^2$ depends on 
$|{\bf{p}}|$ when
${\bf{p}} \to 0$ and therefore the attempt to extract $M_h$ from the low-momentum
data becomes problematic. If, on the other hand, $M_h$ is extracted from the
higher-momentum data where 
$\tilde{E}^2_{\Lambda}({\bf{p}})- {\bf{p}}^2$ does {\it not} depend on 
$|{\bf{p}}|$, then one finds
$\tilde{E}_{\Lambda}(0)< M_h$ with a discrepancy between the two values that 
seems to increase when taking the continuum limit. 

Moreover, 
new data \cite{cea3} show that, for the same lattice
action, by increasing the lattice size one finds
the same $M_h$ at high momenta but smaller and smaller values of
$\tilde{E}_{\Lambda}(0)$.
Namely, by using the same lattice parameters that 
on a $20^4$ lattice give \cite{jansen}
$\tilde{E}_{\Lambda}(0)=0.3912 \pm 0.0012$, one finds \cite{cea3}
$\tilde{E}_{\Lambda}(0)=0.3791 \pm 0.0035$
on a $24^4$ lattice, 
$\tilde{E}_{\Lambda}(0)=0.344\pm 0.008$
on a $32^4$ lattice and
$\tilde{E}_{\Lambda}(0)=0.298\pm 0.015$
on a $40^4$ lattice. Therefore, differently from $M_h$, which is associated with
the higher-momentum part of the energy spectrum, 
the energy-gap in the broken phase is an infrared-sensitive quantity
that, conceivably, vanishes in the infinite-volume limit.

An indication in this sense will be obtained in the next section
by studying the $p \to 0$ limit
of the propagator $G(p)$. The result is very
simple and can be understood by
considering the following equation
\BE
         f^{-1}(x)= 1 +x^2 - g^2 x^2 f(x)
\EE
that bears some analogy to the actual physical situation.
In the  $x \to 0$ limit there are two distinct limiting
behaviours : a) $ f(x) \to 1$ and b) $f(x) \sim {{1}\over{g^2x^2}} \to +\infty$
but only the former solution is recovered with a finite number of iterations
from
\BE
f_0(x)= {{1}\over{1+x^2}}
\EE
for $g^2=0$. 
In the case of $\lambda\Phi^4$ theory, the gapless mode in the broken 
phase corresponds to the b) type of behaviour.

\setcounter{equation}{0}
\section{ The propagator for $p\to 0$ in the broken phase}

Let us now consider a one-component $\lambda\Phi^4$ theory and the problem
of determining the scalar 
propagator of the fluctuating 
field $h(x)=\Phi(x) -\varphi$ in the presence
of a constant background field $\varphi$. 
Ultimately, we shall be interested in the $h-$field
propagator $G(p)$ in the broken symmetry phase, i.e. at the minima of the 
effective potential $V_{\rm eff}(\varphi)$. By defining
\BE
\label{vefft}
{{dV_{\rm eff}}\over{d \varphi}} \equiv J(\varphi)\equiv \varphi T(\varphi^2)
\EE
these are the absolute minima $\varphi=\pm v \neq 0$ where
\BE
\label{1point}         
T(\varphi^2)=0
\EE
and
\BE
\label{second}
   \left. \frac{ d^2 V_{\rm eff}}{d \varphi^2}\right|_{\varphi=\pm v} >0
\EE
Usually, in the broken phase, 
one defines the $h-$field propagator from a Dyson sum of
1PI graphs only, say
\BE
\label{1PI}
G(p)|_{\rm 1PI} \equiv D(p)
\EE
where 
\BE
\label{ident}
D^{-1}(0) \equiv
   {{ d^2 V_{\rm eff} }\over{d \varphi^2}}
\EE
is evaluated at $\varphi =\pm v$. In this way one
neglects the possible role of the
one-particle reducible, zero-momentum tadpole graphs. The reason is  that 
their sum is proportional to the 1-point function, i.e. to 
$J(\varphi)$ in Eq.(\ref{vefft}) 
that vanishes by definition at $\varphi=\pm v$. 
However, the zero-momentum tadpole subgraphs are attached to the other parts
of the  diagrams through zero-momentum propagators. Therefore, their overall
contribution is proportional to
$J(\varphi)G(0)$ that vanishes 
provided $G(0)$ is non-singular at the 
minima. In this respect, neglecting the
tadpole graphs amounts
to {\it assume} the regularity of $G(0)$ at $\varphi=\pm v$ which is certainly
true in a finite-order expansion in powers of 
$J(\varphi)$.
However,  to check the assumption beyond perturbation theory
one has first to control the full propagator in a small region of 
$\varphi-$ values around the minima, by
including all zero-momentum 
tadpole graphs, and finally take the limit $\varphi \to \pm v$.
We observe that the problem of tadpole graphs 
was considered in ref.\cite{munster} where the emphasis was mainly 
to find an efficient way to re-arrange the perturbative expansion.  Here,
we shall attempt a non-perturbative all-order re-summation of the various 
effects to check the regularity of $G(0)$ for $\varphi \to \pm v$.

We shall approach the problem in two steps.
In a first step, we shall consider the contributions to the propagator
that contain all possible insertions of zero-momentum lines on the internal part of the
graphs, i.e. inside 1PI vertices. 
However, at this stage, the external zero-momentum propagators to the sources
maintain their starting value $D(0)$ {\it at} $J=0$.
This approximation gives rise to an auxiliary inverse propagator 
given by 
\BE
\label{aux1}
{G}_{\rm aux}^{-1}(p)= 
D^{-1}(p)
-\varphi z \Gamma_3(p,0,-p)
+{{(\varphi z)^2}\over{2!}}\Gamma_4(0,0,p,-p)
-{{(\varphi z)^3}\over{3!}}\Gamma_5(0,0,0,p,-p)+..
\EE
where
\BE
\label{basic}
             z\equiv T(\varphi^2) D(0)
\EE
represents the basic one-tadpole insertion.
Eq.(\ref{aux1}) can be easily
checked diagrammatically starting from
the tree approximation where 
\BE
\label{tree}
V_{\rm eff}={{1}\over{2}} r \varphi^2 + {{\lambda}\over{4!}} \varphi^4 
\EE
\BE
D^{-1}(p)= p^2 + r + {{\lambda\varphi^2}\over{2}} 
\EE
and
$\Gamma_3(p,0,-p)= \lambda\varphi$, 
$\Gamma_4(0,0,p,-p)= \lambda$
(with all $\Gamma_n$ vanishing for $n>4$). 

Now, by using the relation of the zero-momentum 1PI
vertices with the effective potential at an arbitrary $\varphi$
\BE
       \Gamma_n(0,0,...0)=
        { { d^n V_{\rm eff} }\over {d \varphi^n}}
\EE
we can express the auxiliary
zero-momentum inverse propagator of Eq.(\ref{aux1}) as
\BE
\label{aux2}
G_{\rm aux}^{-1}(0)= 
       \left. \frac{ d^2 V_{\rm eff}}{d \varphi^2} 
\right|_{ {\varphi_{\rm aux}}=\varphi(1 - z) } 
\EE
The second step 
consists in including now
all possible tadpole corrections on each external zero-momentum line in 
(\ref{aux1}). This is independent of the flowing momentum $p$ and
leads to a new infinite hierarchy of Feynman graphs. 
In fact, in a diagrammatic expansion,  a single external zero-momentum
leg gives rise to an infinite number of graphs, each producing
another infinite number of graphs and so on. 
Despite of the apparent complexity of the task, 
the final outcome of this computation can
be cast in a rather simple form, at least on a formal ground.
In fact, we can re-arrange the infinite expansion for the zero-momentum
propagator (all $\Gamma_n$ are evaluated at zero external momenta)
\BE
\label{infinite}
G(0)= D(0) + J\Gamma_3 D^3(0)+ {{3J^2\Gamma^2_3}\over{2}}D^5(0)-
{{\Gamma_4J^2}\over{2}}D^4(0) + {\cal O}(J^3)
\EE
in terms of a modified source
\BE
\label{modsource}
\tilde{J}= J - {{J^2\Gamma_3}\over{2}}D^2(0) +{\cal O}(J^3) \equiv 
\varphi \tilde {T}(\varphi^2)
\EE
in such a way that the full power series expansion for the
exact inverse zero-momentum propagator can be expressed as
\BE
\label{formal}
G^{-1}(0)= 
       \left. \frac{ d^2 V_{\rm eff}}{d \varphi^2} 
\right|_{ {\hat{\varphi}}=\varphi(1 - \tau) } 
\EE
with
\BE
\label{tauu}
            \tau\equiv \tilde{T}(\varphi^2)G(0)
\EE
i.e., as in (\ref{aux2}) with the replacement $z \to \tau$. In the same way, 
Eq.(\ref{aux1}) becomes
\BE
\label{complete}
{G}^{-1}(p)= D^{-1}(p)
- \varphi\tau \Gamma_3(p,0,-p)
+{{(\varphi\tau)^2}\over{2!}}\Gamma_4(0,0,p,-p)
-{{(\varphi\tau)^3}\over{3!}}\Gamma_5(0,0,0,p,-p)+..
\EE
Notice that, in the limit $\varphi \to \pm v$ 
Eq.(\ref{formal}) would be generally considered equivalent to
Eq.(\ref{ident}) that, however, neglects the tadpole graphs altogether. 
As anticipated, this is true provided
$G(0)$ remains non-singular when $J$ and $\tilde{J}$ vanish.
Our main point is that, after resumming the tadpole graphs to all orders, 
there are now multiple solutions for the zero-momentum propagator 
that differ from (\ref{ident}), even when $\varphi \to \pm v$, 
and that would not be present otherwise. 

To study this problem, we shall analyze
Eq.(\ref{formal}) in the case of the
tree approximation Eq.(\ref{tree}). In fact,  the simple idea that the
broken phase has just massive excitations of mass $M_h$ finds its main 
motivation in a tree-level analysis. Moreover, 
we shall assume that when $\varphi \to \pm v$ (and
$ J \to 0$) also $\tilde{J} \to 0$. Assuming
the alternative possibility, i.e. that the full 
$\tilde{J}$ remains non-zero when $\varphi \to \pm v$, would give 
even more drastic results. In fact, in this case, an inverse propagator
as in Eq.(\ref{ident}) would never be recovered from (\ref{formal}), 
even as a particular solution.

By defining the limiting value
$\tau \to \bar{\tau}$ for $\varphi^2 \to v^2$, 
the usual `regular' solution for the inverse propagator (\ref{formal}) 
corresponds to $\bar{\tau}=0$, namely
\BE
\label{regular}
\lim_{\varphi^2 \to v^2} 
{G}^{-1}_{\rm reg}(0)=
{{\lambda v^2}\over{3}}\equiv M^2_h
\EE
so that
\BE
\label{regularp}
{G}^{-1}_{\rm reg}(p)=p^2+ M^2_h
~~~~~~~~~~~~~~~~~~~~{\rm at}~ \varphi=\pm v
\EE 
However, another solution is
\BE
\label{singular}
\lim_{\varphi^2 \to v^2}
{G}^{-1}_{\rm sing}(0)=
{{\lambda v^2}\over{2}}[ \bar{\tau}^2- 2\bar{\tau} + {{2}\over{3}}]=0
\EE
which implies limiting values $ \bar{\tau}=1 \pm {{1}\over{\sqrt{3}}}$ for which
\BE
\label{particular}
{G}^{-1}_{\rm sing}(p)=p^2~~~~~~~~~~~~~~~~~~~~{\rm at}~ \varphi=\pm v
\EE 
Beyond the tree-approximation
finding the singular solution 
$G^{-1}(0)= 0$ at $\varphi=\pm v$ is equivalent to determine that value of 
$\hat{\varphi}^2\equiv v^2(1-\bar{\tau})^2$ where
${{d^2V_{\rm eff}}\over{d \varphi^2}}=0$.
 For instance, in the case of the
Coleman-Weinberg effective potential
\BE
            V_{\rm eff}(\varphi)= {{\lambda^2\varphi^4}\over{256\pi^2}}
(\ln {{\varphi^2}\over{v^2}} -{{1}\over{2}})
\EE
the required values are $\bar{\tau}=1 \pm e^{-1/3}$. In principle, 
such solutions should be found in any
approximation to the effective potential since their existence depends on the
very general assumptions of the broken phase.

Before concluding this section, we shall try to provide a possible
explanation for
the singular zero-momentum behaviour we have pointed out.
To this end, let us introduce the
generating functional for connected Green's functions $W[J]$ and
(for a space-time constant source $J$) 
the associated density $w(J)$
\BE
         W[J]= w(J)\int d^4x
\EE
In this formalism
\BE
\label{phij}
        \phi(J)= {{d w}\over{d J}}
\EE
is just the inverse of
\BE
\label{source1}
                J(\phi)={{dV_{\rm LT} }\over{d\phi}}
\EE
where the effective potential enters as
$V_{\rm eff}(\phi)\equiv V_{\rm LT}(\phi)$, i.e.
the Legendre transform (`LT') of $w(J)$. 
Notice that in this way $V_{\rm LT}(\phi)$ is rigorously
convex downward. For this reason, it is not the same
thing as the usual non-convex (`NC') effective potential
$V_{\rm eff}(\phi)\equiv V_{\rm NC}(\phi)$ we have considered so far. 
Moreover, in the presence of spontaneous symmetry breaking, 
$V_{\rm LT}(\phi)$ is {\it not} 
an infinitely differentiable function of $\phi$ \cite{syma}, differently from
$V_{\rm NC}(\phi)$. In this language, 
$\pm v$ denote the absolute minima of
$V_{\rm NC}(\phi)$ and
Eq.(\ref{ident}) becomes
\BE
\label{newident}
D^{-1}(0)= 
       \left. \frac{ d^2 V_{\rm NC}}{d \phi^2} 
\right|_  {  \phi=\pm v   } 
\EE
Now, let us denote
$J=\hat{J}(\varphi)$ the argument of
 $w(J)$ that corresponds to
determine the full propagator (i.e. including all 
tadpole graphs) in a given background $\varphi$.
We can look for its inversion in the form
\BE
\label{step1}
G^{-1}(0)= 
       \left. \frac{ d^2 V_{\rm LT}}{d \phi^2} 
\right|_  {  \phi=f(\varphi)   } 
\EE
with a suitable $f(\varphi)$. The possibility to recover the same
inverse propagator as in (\ref{newident})
requires that, for $\varphi \to \pm v$, 
the limiting value of $f(\varphi)$ has to
approach one of the absolute minima of the non-convex
effective potential , namely 
\BE
\lim _{|\varphi| \to v} f(\varphi)= \pm v
\EE
Still, (\ref{step1}) is different from (\ref{newident}). In fact, the usual
identifications $V_{\rm LT}(\varphi)=V_{\rm NC}(\pm v)$, 
in the region $-v \leq \varphi \leq v$ enclosed by the absolute
minima of the non-convex approximation, and
                $ V_{\rm LT}(\varphi)=V_{\rm NC}(\varphi)$ 
for $\varphi^2 > v^2$, do not resolve all ambiguities. The identification of
the inverse propagators in Eqs. (\ref{newident}) and (\ref{step1}) amounts 
to a much stronger assumption: 
the derivative in Eq.(\ref{step1}) has to be a left- (or right-)
derivative depending on whether we approach the point $f=-v$ 
( or $f=+ v$). However, this is just a prescription since 
derivatives depend on the chosen path (unless one deals with infinitely 
differentiable functions) and, in general, 
Eq.(\ref{step1}) leads to multiple
solutions for the inverse propagator. Therefore, 
despite one can define a prescription for which
$G^{-1}_{\rm reg}(0)=D^{-1}(0)=M^2_h$ one is also faced with a
$G^{-1}_{\rm sing}(0)=0$, as when approaching the points
$\pm v$ from the inner region 
where the Legendre-transformed potential is flat. 

Now, as discussed in a very transparent way in ref.\cite{ritschel}, 
the difference between $V_{\rm LT}(\phi)$ and
$V_{\rm NC}(\phi)$ amounts to include the
quantum effects of the zero-momentum mode that cannot be
treated as a purely classical background. Namely, one cannot simply
use $V_{\rm NC}(\phi)$ to compute $w(J)$ but has still
to perform one more functional integration in field space on the strength
of the zero-momentum mode. This is the reason why
starting from $w(J)$ one gets $V_{\rm LT}(\phi)$ as the Legendre transform
and {\it not} $V_{\rm NC}(\phi)$. In this respect, our analysis 
suggests that determining the propagator after
this last integration step may finally be
equivalent to include all zero-momentum
tadpole graphs in the classical background.

We conclude this section with the remark that the singular zero-momentum
behaviour we have pointed out does not depend at any stage on the existence
of a continuous symmetry of the classical potential. 
As such, there should be no differences in a spontaneously
broken O(N) theory. Beyond the approximation where
the `Higgs condensate' is treated as a purely
classical background, one has to 
perform one more integration over the zero-momentum mode of the condensed
$\sigma-$field. Therefore, 
all ambiguities in computing the inverse propagator of the
$\sigma-$field through Eq.(\ref{step1}) remain. In this sense, the
existence of gap-less modes of the singlet Higgs field has
nothing to do with the number of field components. 

We only note that in the O(N) theory some additional care is needed to 
re-sum tadpole graphs to all orders in perturbation theory. In fact, 
the tadpole function of the $\sigma-$field 
$T_\sigma(\varphi) \equiv m^2_\pi$ 
plays the role of a mass term for the $\pi$-fields.
By setting $T_\sigma=m^2_\pi=0$ in a straightforward
diagrammatic expansion around $T_\sigma=0$, 
the 2-point function of the $\sigma-$field 
$\Gamma_{\sigma}(p)$ at $p=0$ (the equivalent of our $D^{-1}(0)$)
would become singular due to the massless Goldstone loop. 
However, beyond perturbation theory, $m_\pi=0$ can coexist
with a non-singular $\Gamma_{\sigma}(0)$ \cite{oko}, differently from
the original perturbative analysis of \cite{syma}. In the same 
approximation \cite{oko}, 
the $\pi-$fields decouple from each other and from the $\sigma-$field. 
This result supports the point of view \cite{zeit}
that, beyond perturbation theory, 
the dynamics of `radial' and `angular' degrees of freedom may effectively 
decouple.

\setcounter{equation}{0}
\section{The energy spectrum in the broken phase}

 By choosing the singular solutions corresponding to
${G}^{-1}(0)=0$ at $\varphi=\pm v$, 
one always ends up with a form of
${G}^{-1}(p)$ that vanishes when $p_\mu \to 0$
implying the existence of gap-less modes whose energy 
also vanishes when ${\bf{p}} \to 0$. We can express 
the required relations for the various modes of the cutoff theory
in terms of unknown slope parameters $\eta_\Lambda$ 
\BE
\label{eta}
    \lim_{ {\bf{p}}\to 0 }
 \tilde{E}^2_\Lambda({\bf{p}})=\eta_\Lambda 
{\bf{p}}^2~~~~~~~~~~~~~~~{\rm at}~\varphi=\pm v
\EE
that describe long-wavelength 
collective excitations of the scalar condensate.

Although determining the $\eta_\Lambda$'s  would require the
analytical form of the propagator
at non-zero momenta, one can draw a certain number of
general conclusions. A first observation is that
the two distinct solutions for $\bar{\tau}$ may correspond to
different $\eta_\Lambda$'s meaning that the scalar condensate can support 
different types of oscillations. Think for instance of superfluids where
the velocity of {\it second} sound $c_2$ approaches 
${{c_s}\over{\sqrt{3}}}$ in the limit of zero temperature, $c_s$
being the velocity of (first) sound \cite{pines}. 

A second observation is that the possible limiting
values for the $\eta_\Lambda$'s can easily be guessed by imposing the
requirement of a Lorentz-covariant spectrum in the continuum limit
$\Lambda \to \infty$.
In this case, there are only {\it two} possibilities, namely
\BE
\label{massless}
\eta_\Lambda \to 1
\EE
and/or
\BE
\label{match}
\eta_\Lambda = {\cal O} ( {{M^2_h}\over{\delta^2}} ) \to \infty
\EE
The limit in (\ref{massless}) corresponds to ordinary massless fields, quite
unrelated to the usual massive Higgs particle.
This scenario finds support
in the tree-level approximation Eq.(\ref{particular}).

On the other hand, the alternative (\ref{match})
corresponds to the point of view expressed in the Introduction. This can 
also be understood by noticing that
in the presence of several solutions for the propagator, the physical energy
spectrum is dominated by the lowest excitations for each region of
momenta. Therefore, if there is a massive mode, the form
$\sqrt{ {\bf{p}}^2 +M^2_h}$ has to evolve somehow into
Eq.(\ref{eta}) for $ {\bf{p}}\to 0 $. By denoting
$|{\bf{p}}| \sim \delta\ll M_h$ the typical
range of momenta where the transition 
takes place, this leads to $\eta_\Lambda \delta^2\sim M^2_h$. 
However, exact Lorentz-covariance in the local limit
${{\Lambda}\over{M_h}} \to \infty$ requires that any possible deviation from
$\sqrt{ {\bf{p}}^2 + M^2_h}$ can only reduce to
the zero-measure set $p_\mu=0$, the only
value of the 4-momentum that is left invariant by the
transformations of the Poincar\`e group. Therefore, in the continuum limit
of the broken phase $\eta_\Lambda \to \infty$ meaning that
the form $\sqrt{ {\bf{p}}^2 + M^2_h}$ holds `almost' everywhere, i.e.
with the exception of the point
${\bf{p}}=0$ where one has $\tilde{E}(0)=0$. 
One such example is represented by the toy-model
\BE
\label{toy}
\tilde{E}_{\Lambda}({\bf{p}})=(1- e^{ - {{ |{\bf{p}}| }\over{\delta}} })
\sqrt{ {\bf{p}}^2 + M^2_h}
\EE
with 
\BE
\label{delta}
{{\Lambda}\over{M_h}} =
{{M_h}\over{\delta}} \equiv \sqrt{\eta_\Lambda}
\EE
that leads to 
\BE
\label{toyeq}
\lim_{ {{\Lambda}\over{M_h}} \to \infty}~
{{|\tilde{E}_{\Lambda}({\bf{p}})-\sqrt{ {\bf{p}}^2 + M^2_h}~|}\over
{\sqrt{ {\bf{p}}^2 + M^2_h}}}=
\lim_{ {{\Lambda}\over{M_h}} \to \infty}~
  e^{- {{|{\bf{p}}|}\over{M_h}} {{\Lambda}\over{M_h}} }=0
\EE
provided ${{ |{\bf{p}}| }\over{M_h}} >0$. Nevertheless, for any
$\Lambda$ there is a far infrared region of momenta near ${\bf{p}}=0$ where
$M_h$ and $\tilde{E}_{\Lambda}({\bf{p}})$ differ non trivially. 

This scenario is supported by the experimental example of
superfluid He$^4$. In this case the {\it same} 
spectrum has two different branches: a phonon branch
$\omega_{\rm ph}({\bf{p}})= c_s|{\bf{p}}|$ describing the spectrum for
$ {\bf{p}}\to 0 $ and a roton branch
$\omega_{\rm rot}({\bf{p}}) = \Delta + {{|{\bf{p}}|^2 }\over{ 2\mu}}$ 
starting at momenta $|{\bf{p}}| > p_0$. With this analogy, 
(\ref{match}) would represent
the matching at $|{\bf{p}}| \sim \delta \ll M_h $
between a `phonon'
with sound velocity $c_s=\sqrt{\eta_\Lambda}$ and a `Lorentz-covariant roton' with
$\Delta = \mu \equiv M_h$.

\setcounter{equation}{0}
\section{Summary and outlook}
\par In this Letter, we have presented several arguments that, 
quite independently of the Goldstone phenomenon, 
suggest the existence of gap-less modes of the scalar condensate in the
broken phase of $\lambda\Phi^4$ theories. Although the argument 
can be purely
diagrammatic, this result will be better understood
by representing spontaneous symmetry breaking along the
lines presented in the Introduction, i.e.
as a real condensation process of physical quanta \cite{mech}.
In this way, one will also understand the puzzling lattice data of
ref.\cite{cea3}. These show that 
the energy-gap in the broken phase of a one-component $\lambda\Phi^4$ theory
is {\it not} the `Higgs mass' $M_h$ but an
infrared-sensitive quantity that becomes smaller and  smaller by increasing
the lattice size.

Now, it is well known that
condensed-matter systems can support long-range forces
even with elementary constituents that only have short-range 2-body 
interactions. For this reason, it should not be surprising that
a gap-less collective mode in the 
condensed phase of $\lambda\Phi^4$ theories
will give rise to a long-range potential. 
The phenomenological aspects of this analysis and a possible wider
theoretical framework will be presented elsewhere.
\vskip 5 pt
{\bf Acknowledgements}~~I thank Dario Zappala' for pointing out a mistake in
the original version of this paper.

\end{document}